\documentclass[lineno]{aero}
\let\linenumbershook\empty
\usepackage{algorithm2e}

\ADsetup{
title    = {Differential Corrections Algorithm for Initial Orbit Determination in the Cislunar Region using Angle-Only Measurements},
author   = {Seur Gi Jo$^{1}$\cor{}, Brian Baker-McEvilly$^{1}$, and David Canales$^{1}$ 
           },
address  = {
  1. Embry-Riddle Aeronautical University, 1 Aerospace Boulevard, Daytona Beach, FL 32114, USA
},
email    = {jos2@my.erau.edu},
abstract = {
As space traffic continues to increase in the cislunar region, accurately determining the trajectories of objects operating within this domain becomes critical. However, due to the combined gravitational influences of the Earth and Moon, orbital dynamics in this region are highly nonlinear and often exhibit chaotic behavior, posing significant challenges for trajectory determination. Many existing methods attempt to address this complexity using machine learning models, advanced optimization techniques, or sensors that directly measure distance to the target, approaches that often increase computational burden and system complexity. This work presents a novel initial orbit determination (IOD) algorithm for cislunar objects based solely on three angle-only measurements taken at three discrete times. The core methodology builds upon a differential corrections framework which iteratively refines the target's trajectory to satisfy line-of-sight constraints. Numerical simulations demonstrate that the algorithm enables accurate IOD using minimal observational data, making it particularly suited for onboard implementation in resource-constrained cislunar missions.
  },
keywords = {
  Initial orbit determination \sep Circular restricted three-body problem \sep Differential corrections \sep Angle-only navigation \sep Cislunar region \\
},
}

\begin{document}

\maketitle  


\Nomenclature

\begin{center}
\begin{tabular}{|P{.45\linewidth}|P{.45\linewidth}|}
\hline
$DF$ & Jacobian of constraints with respect to free variables \\
\hline
$\vec{F}$ & Constraint vector \\
\hline
$l^*$ & Characteristic length (m)\\
\hline
$m^*$ & Characteristic mass (kg)\\
\hline
$\vec{r}$ & Position vector \\
\hline
$t^*$ & Characteristic time (s)\\
\hline
$U^*$ & Pseudo-potential function \\
\hline
$\vec{v}$ & Velocity vector \\
\hline
$\vec{x}$ & State vector \\
\hline
$\alpha$ & Range between an observer and a target \\
\hline
$\vec{\chi}$ & Vector of free variables \\
\hline
$\Phi$ & State transition matrix \\
\hline
$\hat{\rho}$ &  Relative unit vector from an observer to a target  \\
\hline
\end{tabular}
\end{center}

\section{Introduction}
Recent advancements in the space industry have yielded numerous missions beyond Earth's orbit \cite{BAKERMCEVILLY2024101019, baker2024performance}. As the number and diversity of these missions grow, it becomes increasingly important to monitor and analyze the motion of objects operating within the cislunar environment. In particular, accurate initial orbit determination (IOD) is critical for many space operations, including spacecraft rendezvous and the identification of potential collision threats. However, the cislunar region presents several unique challenges compared to near-Earth orbits. In terms of observation challenges, observation distances are significantly larger, radar sensors used to monitor objects near the Earth cannot reach far into cislunar space, and long-range optical sensor infrastructure is limited and over-tasked. Dynamical challenges also persist, as the gravitational influence of both the Earth and Moon shall be considered simultaneously, leading to complex and often highly nonlinear trajectories characteristic \cite{seurgi2025knot}. These dynamics, governed by multi-body interactions, render classical IOD techniques such as the Gauss or Laplace methods either ineffective or unreliable in this regime \cite{escobal1970methods, gooding1993new, fujimoto2012correlation, vallado}.

To address these challenges, recent research has proposed various alternative approaches tailored to angle-only IOD in the cislunar environment. Smego and Christian introduced a dynamic triangulation method that generates an initial state estimate from line-of-sight (LOS) measurements taken by a single observer at multiple epochs \cite{smego2024}. Such an approach requires continuous tracking of the target and involves trade-offs between observation frequency and duration. Zuehlke et al. proposed two approaches to angle-only IOD: one using a particle swarm optimizer to minimize measurement residuals, and another leveraging a semi-analytical formulation based on linearized relative motion and constraint equations \cite{zuehlke2022,zuehlke2024}. While both methods effectively handle nonlinear dynamics, they suffer from high computational cost or sensitivity to initial guesses.
Other strategies have incorporated alternative measurement types to improve angle-only IOD. Waggoner et al. introduced a method that augments angular observations with multi-receiver Doppler ratios, enabling range observability without direct range sensors \cite{waggoner2024}. However, the approach requires cooperative targets and a synchronized receiver network, which may not be feasible in resource-constrained missions. Liu et al. developed a perturbed-Laplace method that extends classical Laplace’s formulation to account for Earth-Moon gravitational perturbations and analyzes the multiplicity of resulting solutions \cite{liu2025}. While theoretically comprehensive, the method relies on higher-order polynomial fitting and is highly sensitive to observation noise and coplanar geometries, which may lead to missed or spurious solutions in practice. Notably, Heidrich and Holzinger developed a sparse grid collocation method tailored for angle-only IOD with limited and temporally sparse observations \cite{heidrich2025}. Their approach enables orbit estimation under challenging observation schedules by efficiently sampling the admissible state space. Such a method incurs significant computational cost due to high-dimensional grid evaluations and is sensitive to observation geometry and noise. Dinh et al. introduced a sensor- and measurement-centric admissible region method that employs topocentric intersection theory to constrain the initial range and range-rate space using only angle and angle-rate measurements \cite{dinh2024}. This approach enables a geometric visualization of feasible initial states without requiring direct range observations. Nevertheless, it is computationally intensive due to the need to propagate a dense field of virtual objects, and it suffers from numerical sensitivity when the time between observations is short. Data-driven techniques have also gained traction, including an adaptive machine learning framework that leverages historical data to improve estimation performance \cite{ojeda2024}. In summary, many of the proposed approaches rely on collocation techniques, optimization algorithms, or machine learning methods that are inherently complex and may exceed the computational capabilities of small spacecraft.

In this paper, a novel angle-only IOD approach is proposed for the cislunar region using the circular restricted three-body problem (CR3BP). The method assumes an observer with a known position in the cislunar region, viewing a target at multiple distinct epochs. Using only LOS measurements at each epoch, a differential corrections scheme is employed to estimate the target’s velocity and relative distance~\cite{Garcia2021}. This approach enables spacecraft operating in the cislunar environment to autonomously determine the target's state using a relatively simple and lightweight algorithm. While the method shares the use of the CR3BP model and state transition matrix (STM) with prior efforts such as Zuehlke et al.~\cite{zuehlke2024}, it differs in both scope and application. The present method addresses absolute orbit determination from a known observer using angle-only measurements, whereas Zuehlke’s approach focuses on relative orbit determination based on semi-analytical constraints tailored to near-target scenarios. This distinction makes the proposed method particularly useful when the observer and target are not in close proximity.

This paper is organized as follows: Section~\ref{sec:CR3BP} introduces the CR3BP dynamical model and the associated state transition matrix. Section~\ref{sec:IOD} details the proposed angle-only IOD method using relative unit vectors from the observer to the target. Section~\ref{sec:Simulation} presents numerical simulations under four representative scenarios and a discussion of the results. Finally, Section~\ref{sec:Conclusion} offers concluding remarks and potential directions for future work.

\section{Dynamical Model} \label{sec:CR3BP}
\begin{figure}[!h]
\centering\includegraphics[width=0.7\linewidth]{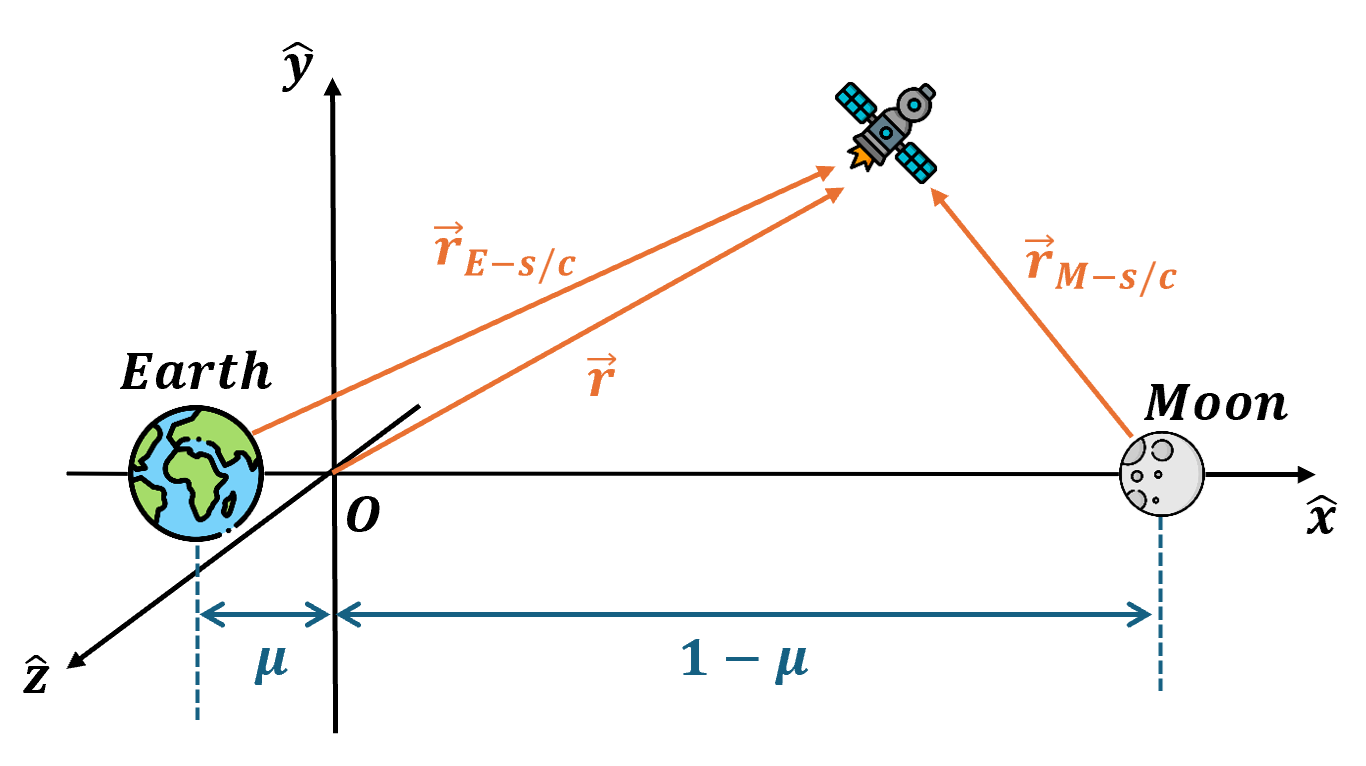}
	\caption{CR3BP schematic}
	\label{fig:CR3BP}
\end{figure}

In order to execute the proposed IOD algorithm, a dynamical model that accurately represents the cislunar region is needed. In this work, the CR3BP is adopted, as it is a well-known and thoroughly studied model. This section provides a brief overview of the CR3BP and describes the STM, which is utilized later in the IOD algorithm. In the CR3BP, the two primaries orbit their mutual barycenter in circular orbits within the same orbit plane. The positions of the primaries are defined using the mass ratio of the system:
\begin{align}
\label{eq:MassParameter}
\mu=\frac{m_{M}}{m_{E}+m_{M}},
\end{align}
where $m$ is the mass and the subscripts $E$ and $M$ refer to the Earth and Moon, respectively. In the Earth-Moon system, $\mu$ is approximately $0.01215$. A third body, such as a spacecraft, has negligible mass and thus exerts no gravitational influence on the two primaries. Although the Moon’s actual orbital eccentricity is approximately $0.05488$, the CR3BP suffices as an accurate fidelity model for describing the motion of objects in the cislunar environment. The equations of motion are formulated in the Earth-Moon rotating frame, where the origin is at the barycenter of the Earth-Moon system, the $\hat{x}$-axis points from the Earth to the Moon, the $\hat{z}$-axis out of the orbital plane along the angular momentum vector of the system, and the $\hat{y}$-axis completes the right-handed coordinate system. 
The governing equations of motion in this rotating frame are expressed as follows:
\begin{equation}
    \ddot{x}-2\dot{y}=\frac{\partial U^*}{\partial x}, \qquad \ddot{y}+2\dot{x}=\frac{\partial U^*}{\partial y}, \qquad \ddot{z}=\frac{\partial U^*}{\partial z},
\end{equation}
where $U^*$ is the pseudo-potential function defined as:
\begin{equation} \label{eqn:pseudo-potential}
    U^* = \frac{1-\mu}{\|\vec{r}_{E-s/c}\|}+\frac{\mu}{\|\vec{r}_{M-s/c}\|}+\frac{1}{2}(x^2 + y^2),
\end{equation}
with $\vec{r}_{E-s/c}$ and $\vec{r}_{M-s/c}$ denoting the position vectors from the Earth and Moon to the spacecraft, respectively. To simplify the equations of motion, non-dimensional units are typically used. The characteristic mass $m^*$ is the total mass of the two primaries, the characteristic length $l^*$ is the Earth-Moon distance, and the characteristic time $t^*$ is given by $t^* = \sqrt{\frac{{(l^*)}^3}{G m^*}}$, where $G$ is the gravitational constant. A schematic of the Earth-Moon CR3BP is provided in Fig.~\ref{fig:CR3BP}.

\subsection{State Transition Matrix} \label{sec:STM}
To obtain the STM for this nonlinear system, the equations are linearized via a Taylor series expansion:
\begin{equation} \label{eq:taylorseries}
f(\vec{x}) - f(\vec{x}_0) = f'(\vec{x}_0)(\vec{x}-\vec{x}_0) + \frac{f''(\vec{x}_0)}{2!}(\vec{x}-\vec{x}_0)^2 + \cdots,
\end{equation}
where the state vector is defined as $\vec{x}_0 = \left[ \vec{r}^\top, \vec{v}^\top \right]^{\top} = \left[ x, y, z, \dot{x}, \dot{y}, \dot{z} \right]^{\top}$ and $f(\vec{x}_0) = \left[ \dot{\vec{r}}^\top, \dot{\vec{v}}^\top \right]^{\top}$. A perturbed state $\vec{x}$ is expressed as $\vec{x} = \vec{x}_0 + \delta\vec{x} = \left[ x+\xi, y+\eta, z+\zeta, \dot{x}+\dot{\xi}, \dot{y}+\dot{\eta}, \dot{z}+\dot{\zeta} \right]^{\top}$. Neglecting higher-order terms yields the linearized system:
\begin{equation} \label{eq:HOT}
    \delta\dot{\vec{x}} = f'(\vec{x}_0)\delta\vec{x},
\end{equation}
where $f'(\vec{x}_0)$ is the Jacobian matrix, $A(\vec{x}_0)\in \mathbb{R}^{6\times6}$ \cite{jo2024analysis}. For the CR3BP, Eq. \eqref{eq:HOT} is explicitly written as
\begin{equation} \label{eq:Jacobian}
\begin{split}
\underbrace{\begin{bmatrix}
    \dot{\xi} \\
    \dot{\eta} \\
    \dot{\zeta} \\
    \ddot{\xi} \\
    \ddot{\eta} \\
    \ddot{\zeta}
\end{bmatrix}}_{\delta\dot{\vec{x}}} = \underbrace{\begin{bmatrix}
0 & 0 & 0 & 1 & 0 & 0 \\
0 & 0 & 0 & 0 & 1 & 0 \\
0 & 0 & 0 & 0 & 0 & 1 \\
U_{xx}^* & U_{xy}^* & U_{xz}^* & 0 & 2 & 0 \\
U_{xy}^* & U_{yy}^* & U_{yz}^* & -2 & 0 & 0 \\
U_{xz}^* & U_{yz}^* & U_{zz}^* & 0 & 0 & 0
\end{bmatrix}}_{A(\vec{x}_0)}\underbrace{\begin{bmatrix}
    \xi \\
    \eta \\
    \zeta \\
    \dot{\xi} \\
    \dot{\eta} \\
    \dot{\zeta}
\end{bmatrix}}_{\delta\vec{x}},
\end{split}
\end{equation}
where the second derivatives of the pseudo-potential are
\begin{equation} \label{eq:Jacobian3}
\begin{split}
U_{xx}^* =& 1 - \frac{1-\mu}{r_{E-s/c}^3} + \frac{3(1-\mu)(x + \mu)^2}{r_{E-s/c}^5} - \frac{\mu}{r_{M-s/c}^3} + \frac{3\mu(x - 1 + \mu)^2}{r_{M-s/c}^5},\\
U_{yy}^* =& 1 - \frac{1-\mu}{r_{E-s/c}^3} + \frac{3(1-\mu)y^2}{r_{E-s/c}^5} - \frac{\mu}{r_{M-s/c}^3} + \frac{3\mu y^2}{r_{M-s/c}^5}, \\
U_{zz}^* =& -\frac{1-\mu}{r_{E-s/c}^3} + \frac{3(1-\mu)z^2}{r_{E-s/c}^5} - \frac{\mu}{r_{M-s/c}^3} + \frac{3\mu z^2}{r_{M-s/c}^5}, \\
U_{xy}^* =& \frac{3(1-\mu)(x+\mu)y}{r_{E-s/c}^5} + \frac{3\mu(x - 1 + \mu)y}{r_{M-s/c}^5}, \\
U_{yz}^* =& \frac{3(1-\mu)yz}{r_{E-s/c}^5} + \frac{3\mu yz}{r_{M-s/c}^5}, \\
U_{xz}^* =& \frac{3(1-\mu)(x+\mu)z}{r_{E-s/c}^5} + \frac{3\mu(x - 1 + \mu)z}{r_{M-s/c}^5}.
\end{split}
\end{equation}
The STM, $\Phi(t_k, t_0) \in \mathbb{R}^{6\times6}$, is then obtained by numerically integrating:
\begin{equation}
    \dot{\Phi}(t_k, t_0) = A\left(\vec{x}(t_k)\right) \Phi(t_k, t_0),
\end{equation}
with the initial condition $\Phi(t_0, t_0) = I_6$, where $I_6\in \mathbb{R}^{6\times6}$ is the identity matrix, $t_0$ is the initial time of propagation, and $t_k$ is the $k$-th time. The STM thus bridges the dynamics between the two epochs. The STM satisfies the composition property: $\Phi(t_2, t_1)\Phi(t_1, t_0) = \Phi(t_2, t_0)$, and the mapping property: $\delta\vec{x}(t_f)=\Phi(t_f, t_0)\delta\vec{x}(t_0)$. Physically, the STM is a powerful tool that maps perturbations from the initial state, $\delta\vec{x}(t_0)$, to perturbations in the final state, $\delta\vec{x}(t_f)$, providing insight into the sensitivity of a trajectory's evolution. It is also worth noting that the CR3BP equations of motion are time-reversible, which enables backward-time numerical integration for STM computation. This property is leveraged in IOD and trajectory design algorithms where reverse-time propagation is beneficial.

\section{Angle-Only Initial Orbit Determination Algorithm} \label{sec:IOD}
This section presents a differential corrections–based IOD algorithm designed for angle-only measurements in the cislunar environment. The method, described in Section~\ref{sec:ThreeEpoch}, estimates the target's initial state using three relative unit vectors taken at distinct epochs, assuming known observer positions and initial guesses of the target range at each epoch. Although the problem is well-constrained under this formulation, the limited nature of angle-only information may still admit multiple feasible solutions depending on the initial guess. To address this, Section~\ref{sec:Initialization} outlines a practical initialization strategy to improve convergence reliability. Finally, Section~\ref{sec:Summary} summarizes the overall algorithm and highlights key considerations for implementation.

\subsection{Differential Correction–Based Method} \label{sec:ThreeEpoch}
\begin{figure}[b!]
    \hfill{}
    \centering
    \includegraphics[width=0.45\linewidth]{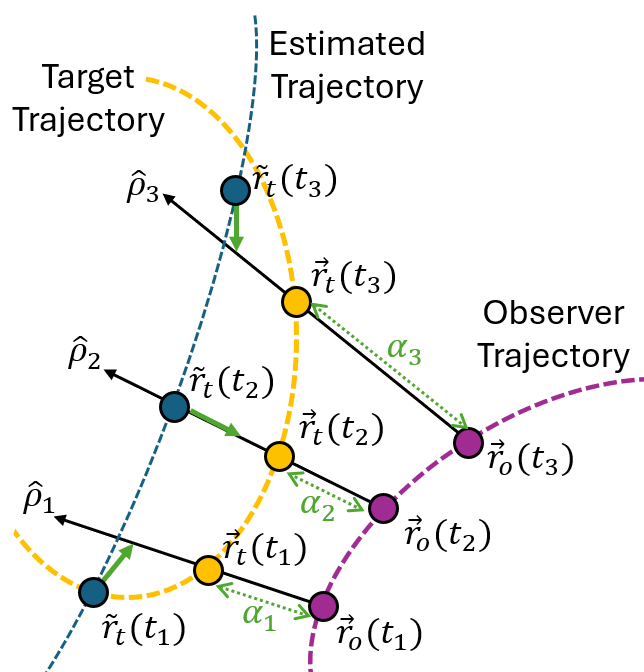}\hfill{}
    \caption{\label{fig:IOD}Schematic of the IOD system.}
\end{figure}

The proposed IOD algorithm offers a novel method for determining trajectories in the cislunar region requiring three angle-only measurements and an initial guess of the range from the observer to the target. The algorithm assumes that the position of the observer is known during each observation, and the observer completes three measurements across three distinct times, measuring sets of relative angles to the target. To maintain generality, the sets of angles are converted into relative unit vectors representing the target's position relative to the observer. The IOD algorithm aims to directly estimate the relative position and velocity of the target object at the time of the interior observation, i.e., at the instant of the second measurement. The measurements are supplied at three times, denoted as $t_1$, $t_2$, and $t_3$, containing relative angles that are transformed into relative unit vectors, denoted as $\hat{\rho}_1$, $\hat{\rho}_2$, and $\hat{\rho}_3$, respectively for each time. The transformation from angles to relative unit vectors depends on the type of angles measured, such as right ascension and declination or azimuth and elevation, and involves applying the appropriate rotation matrices. Regardless of the angle type, all measurements are ultimately converted into relative unit vectors. For example, when the observed angles are azimuth ($a$) and elevation ($h$), the corresponding relative unit vector in the Earth–Moon rotating frame is given by:
\begin{equation}
    \hat{\rho} = R
\begin{bmatrix}
\cos h \cos a \\
\cos h \sin a \\
\sin h
\end{bmatrix},
\end{equation}
where $R$ is the rotation matrix that transforms vectors from the observer's body frame to the Earth–Moon rotating frame. Additionally, the observer's positions ($\vec{r}_o$) are known at each measurement time. The position of the target at each observation time corresponds to
\begin{equation}
\begin{split}
\vec{r}_t(t_1) = \vec{r}_o(t_1) + \alpha_1 \hat{\rho}_1,\qquad
\vec{r}_t(t_2) = \vec{r}_o(t_2) + \alpha_2 \hat{\rho}_2,\qquad
\vec{r}_t(t_3) = \vec{r}_o(t_3) + \alpha_3 \hat{\rho}_3, 
\end{split}
\label{eq:PosRel}
\end{equation}
where $\alpha_1$, $\alpha_2$, and $\alpha_3$ represent the range between the observer and the target at each observation time, which are unknown. A schematic of this system is provided in Fig.~\ref{fig:IOD}.
The IOD algorithm estimates the state at $t_2$, 
\begin{equation}
\label{eq:iod1}
\vec{x}(t_2) = \begin{bmatrix}
    \vec{r}_o(t_2) + \alpha_2 \hat{\rho}_2 \\ \vec{v}(t_2) 
\end{bmatrix} \quad\in\mathbb{R}^{6},
\end{equation}
where $\vec{v}(t_2)$ is the velocity of the interior point and is also unknown. Thus, in the described system, there are a total of six variables that are unknown and of interest: $\alpha_1$, $\alpha_2$, $\alpha_3$, and $\vec{v}(t_2)$. Since these points are all along a common trajectory, it is true that the interior point propagated forward in time to $t_3$ shall converge to the position $\vec{r}_t(t_3)$, and the interior point propagated backward in time to $t_1$ shall converge to the position $\vec{r}_t(t_1)$. Therefore, two equations are formulated, totaling six individual constraints:
\begin{eqnarray}
\vec{F}_1 = \vec{r}_t(t_1) - \tilde{r}_t(t_1) = \delta \vec{r}_t\left(t_1 \right) =  \vec{r}_o(t_1)+\alpha_1 \rho_1 - \tilde{r}_t(t_1) = 0_{3\times1}, \label{eq:F1} \\
\vec{F}_2 = \vec{r}_t(t_3) - \tilde{r}_t(t_3) = \delta \vec{r}_t\left(t_3 \right) = \vec{r}_o(t_3)+\alpha_3 \rho_3 - \tilde{r}_t(t_3) =  0_{3\times1}, \label{eq:F2}
\end{eqnarray}
where $\vec{r}_t(t)$ denotes the true target positions, while $\tilde{r}_t(t_3/t_1)$ represents the interior measurement's state propagated forwards/backwards to time $t_3/t_1$, respectively. To satisfy the six constraints, the six unknown quantities, $\alpha_1$, $\alpha_2$, $\alpha_3$, and $\vec{v}(t_2)$, are treated as free variables in a differential corrections scheme. For a fully constrained problem, the update equation for the free variables is defined as:
\begin{equation}
\label{eq:iod1}
\vec{\chi}^{(i+1)} = \vec{\chi}^{(i)} - {DF\left(\vec{\chi}^{(i)} \right)}^{-1} \vec{F}\left(\vec{\chi}^{(i)}\right),
\end{equation}
where $\vec{\chi}^{(i)} = {\left[ \alpha_1^{(i)}, \alpha_2^{(i)}, \alpha_3^{(i)}, {\vec{v}_t^{(i)}(t_2)}^{\top} \right]}^{\top}$ is the vector of free variables, $\vec{F}$ is the vector of constraints, and $DF\left(\vec{\chi}^{(i)} \right) = \frac{\partial \vec{F}_n}{\partial \vec{\chi}_m}\rvert_{\vec{\chi}^{(i)}}$ is the Jacobian of $n$ constraints with respect to $m$ free variables evaluated at $\vec{\chi}^{(i)}$, where the superscript $(i)$ denotes the value of the variable at the $i$-th iteration of the differential corrections process. In this case, the problem is well-constrained, consisting of six constraint equations and six free variables, which allows for the computation of a unique numerical solution for a given initial guess. However, there is no apparent direct relationship between the free variables and the constraints. Therefore, to determine $DF\left(\vec{\chi}^{(i)} \right)$ analytically for the given free variables, further manipulation of the constraints is needed. Recall the STM mapping property, in which a state at time $t_0$ may be mapped to the state at time $t_f$ using $\delta\vec{x}(t_f)=\Phi(t_f, t_0)\delta\vec{x}(t_0)$. Further broken into components,
\begin{equation}
    \begin{bmatrix}\delta\vec{r}_t(t_f)\\ \delta\vec{v}_t(t_f)
\end{bmatrix}=\begin{bmatrix}
    \Phi_{11}(t_f, t_0) & \Phi_{12}(t_f, t_0) \\ \Phi_{21}(t_f, t_0) & \Phi_{22}(t_f, t_0) 
\end{bmatrix} \begin{bmatrix}\delta\vec{r}_t(t_0)\\ \delta\vec{v}_t(t_0)
\end{bmatrix},
\end{equation}
is obtained, where the change in the final position is mapped to the change in initial conditions by: 
\begin{equation} \label{eq:STMConst}
\delta\vec{r}_t(t_f) = 
    \Phi_{11}(t_f, t_0) \delta\vec{r}_t(t_0) +\Phi_{12}(t_f, t_0) \delta\vec{v}_t(t_0) .
\end{equation}
To apply the STM relation in Eq.\eqref{eq:STMConst} to the IOD formulation, the initial time $t_0$ is set to the second observation time $t_2$, and the final time $t_f$ is set to either $t_1$ or $t_3$, yielding the following expressions:
\begin{equation} \label{eq:eqset1}
\begin{split}
\delta\vec{r}_t(t_1) &= \Phi_{11}^-(t_1, t_2) \delta\vec{r}_t(t_2) +\Phi_{12}^-(t_1, t_2) \delta\vec{v}_t(t_2), \\
\delta\vec{r}_t(t_3) &= \Phi_{11}^+(t_3, t_2) \delta\vec{r}_t(t_2) +\Phi_{12}^+(t_3, t_2) \delta\vec{v}_t(t_2).
\end{split}
\end{equation}
Here, the superscripts $+$ and $-$ indicate forward and backward propagation from the interior point at $t_2$, respectively. Both $\delta\vec{r}_t(t_2)$ and $\delta\vec{v}_t(t_2)$ represent the differences in the target state across successive iterations:
\begin{equation} \label{eq:eqset2}
\begin{split}
\delta\vec{r}_t(t_2) &= (\alpha_2^{(i+1)} - \alpha_2^{(i)}) \hat{\rho}_2, \\
\delta\vec{v}_t(t_2) &= \vec{v}_t^{(i+1)}(t_2) - \vec{v}_t^{(i)}(t_2).
\end{split}
\end{equation}
Similarly, $\delta\vec{r}_t(t_1)$ and $\delta\vec{r}_t(t_3)$ are written based on the constraint functions in Eqs. \eqref{eq:F1} and \eqref{eq:F2} as:
\begin{equation} \label{eq:eqset3}
\begin{split}
\delta\vec{r}_t(t_1) &= \vec{r}_o(t_1)+\alpha_1^{(i)} \rho_1 - \tilde{r}_t(t_1), \\
\delta\vec{r}_t(t_3) &= \vec{r}_o(t_3)+\alpha_3^{(i)} \rho_3 - \tilde{r}_t(t_3).
\end{split}
\end{equation}
By substituting Eqs.\eqref{eq:eqset2} and \eqref{eq:eqset3} into Eq.\eqref{eq:eqset1}, the constraint functions are rewritten as the following augmented equations:
\begin{align}
\vec{F}_1&=\vec{r}_o(t_1) + \alpha_1^{(i)} \hat{\rho}_1  - \tilde{r}_t(t_1) \nonumber\\
&= \Phi^-_{11}(t_1, t_2)\left(\alpha_2^{(i+1)} - \alpha_2^{(i)}\right) \hat{\rho}_2 + \Phi^-_{12}(t_1, t_2)\left(\vec{v}_t^{(i+1)}(t_2) - \vec{v}_t^{(i)}(t_2)\right)=0_{3\times1}, \label{eq:F1Ex} \\[1em]
\vec{F}_2&=\vec{r}_o(t_3) + \alpha_3^{(i)} \hat{\rho}_3  - \tilde{r}_t(t_3)\nonumber \\
&= \Phi^+_{11}(t_3, t_2)\left(\alpha_2^{(i+1)} - \alpha_2^{(i)}\right) \hat{\rho}_2 + \Phi^+_{12}(t_3, t_2)\left(\vec{v}_t^{(i+1)}(t_2) - \vec{v}_t^{(i)}(t_2)\right)=0_{3\times1}. \label{eq:F2Ex}
\end{align}
Next, by collecting all known terms from the $i$-th iteration and unknown terms from the $(i+1)$-th iteration, Eqs.~\eqref{eq:F1Ex} and \eqref{eq:F2Ex} are reorganized into the following matrix form:
\begin{equation}\label{eq:matrixform}
\underbrace{
\begin{bmatrix}
\!\alpha_1^{(i+1)}\!\\
\!\alpha_2^{(i+1)}\!\\
\!\alpha_3^{(i+1)}\!\\
\! \vec{v}_t^{(i+1)}\!(t_2)\!
\end{bmatrix}
}_{\vec{\chi}^{(i+1)}}=
\underbrace{
\begin{bmatrix}
  \!  \alpha_1^{(i)} \!\\
  \!  \alpha_2^{(i)} \!\\
  \!  \alpha_3^{(i)} \!\\
  \!  \vec{v}_t^{(i)}\!(t_2)\! 
\end{bmatrix}
}_{\vec{\chi}^{(i)}}-
\underbrace{
\begin{bmatrix} 
\!\hat{\rho}_1 \!&\! -\Phi^-_{11}(t_1, t_2) \hat{\rho}_2 \!&\! 0_{3\!\times\!1} \!&\! -\Phi^-_{12}(t_1, t_2) \!\\
\!0_{3\!\times\!1} \!&\! -\Phi^+_{11}(t_3, t_2) \hat{\rho}_2 \!&\! \hat{\rho}_3 \!&\! -\Phi^+_{12}(t_3, t_2) \!
\end{bmatrix}^{-1}
}_{DF(\vec{\chi}^{(i)})^{-1}}
\underbrace{
\begin{bmatrix}
   \! \vec{r}_o(t_1)\! +\! \alpha_1^{(i)} \hat{\rho}_1  \!- \! \tilde{r}_t(t_1)\!\\
   \! \vec{r}_o(t_3) \!+\! \alpha_3^{(i)} \hat{\rho}_3  \!-\!  \tilde{r}_t(t_3)\!
\end{bmatrix}
}_{\vec{F}(\vec{\chi}^{(i)})}.
\end{equation}
Equation~\eqref{eq:matrixform} represents the explicit update rule in the form of Eq.~\eqref{eq:iod1}, completing the formulation of the differential corrections algorithm. Overall, the algorithm utilizes differential corrections to satisfy a set of position constraints that ensures the propagation of the interior measurement follows along the corresponding relative unit vector found from the measurements taken at the initial and final times.

\subsection{Initial Guess Selection} \label{sec:Initialization}
Initial guesses of the free variables, i.e., the ranges at each measurement and the interior measurement's velocity, are required to initialize the proposed IOD algorithm. Due to the nonlinear nature of the differential corrections process and the inherent ambiguity in angle-only measurements, the final converged solution is highly sensitive to the choice of initial guesses. Different initializations may lead to distinct local solutions that satisfy the measurement constraints within tolerance. As such, careful selection of the initial range values plays a crucial role in ensuring convergence to a physically meaningful solution. Among these, the range guesses at each epoch ($\alpha_1$, $\alpha_2$, and $\alpha_3$) often rely on contextual information. When the apparent brightness (luminosity), physical size of the target (which may be assumed to be with the commonly accepted cannonball model), and the characteristics of the onboard camera are known, the range $\alpha$ is estimated using photometric relationships. One such model expresses the reflected irradiance as a function of the target’s reflective and geometric properties~\cite{BAKERMCEVILLY2024101019, Frueh2019AMOS, Frueh2021CislunarSSA}:
\begin{equation} \label{eq:luminosity}
\alpha = \sqrt{\frac{2}{3}\frac{I_{\text{sun}}}{I_t}  \frac{C_d}{\pi^2}d^2 \left( \sin\theta + (\pi - \theta)\cos\theta \right)},
\end{equation}
where $I_t$ denotes the irradiance reflected from the target, $I_{\text{sun}}$ represents the solar reference irradiance, $C_d$ is the Lambertian diffuse reflection coefficient, $d$ is the target radius, and $\theta$ is the phase angle between the observer and the Sun at the target location. This equation provides a coarse but informative estimate of the target’s relative distance ($\alpha$), which is used as an initial guess. Once the range values are estimated, they define the initial guess for the target’s velocity at the interior measurement time. This estimate follows from central differencing:
\begin{equation}
\vec{v}_t(t_2) = \frac{\vec{r}_t(t_3) - \vec{r}_t(t_1)}{t_3 - t_1} = \frac{\left( \vec{r}_o(t_3)  + \alpha_3 \hat{\rho}_3 \right) - \left( \vec{r}_o(t_1) + \alpha_1 \hat{\rho}_1 \right)}{t_3 - t_1},
\end{equation}
which relates the velocity at $t_2$ to the initial guesses for $\alpha_1$ and $\alpha_3$. This approach removes the need for a separate velocity guess and directly links the velocity estimate to the selected range values.

The IOD method proposed in this paper may converge to different solutions that all satisfy the LOS constraints, particularly in cases where multiple trajectories are mathematically admissible. In such situations, the final outcome depends on the initial range guesses. If sufficient a priori information is unavailable to guide the selection of these initial guesses, it becomes difficult to identify which of the resulting solutions corresponds to the true target state. To verify whether the solution obtained from a given initial guess reflects the actual trajectory, an additional fourth angle-only measurement can be incorporated. Specifically, after performing IOD using ${t_1, t_2, t_3}$ and obtaining a candidate solution, the estimated range values $\alpha_2$ and $\alpha_3$ are reused as initial guesses for a second IOD process using measurements at ${t_2, t_3, t_4}$. If the results from the second estimation are consistent with the original solution, the candidate can be considered valid. This disambiguation strategy is simple and lightweight, requiring neither range measurements nor cooperative targets, making it well-suited for autonomous onboard applications. An example implementation of this approach is demonstrated in the simulation results in Section~\ref{sec:Simulation}.

\subsection{Summary and Remarks} \label{sec:Summary}

\begin{figure}[!htb]
    \centering
    \includegraphics[width=1\linewidth]{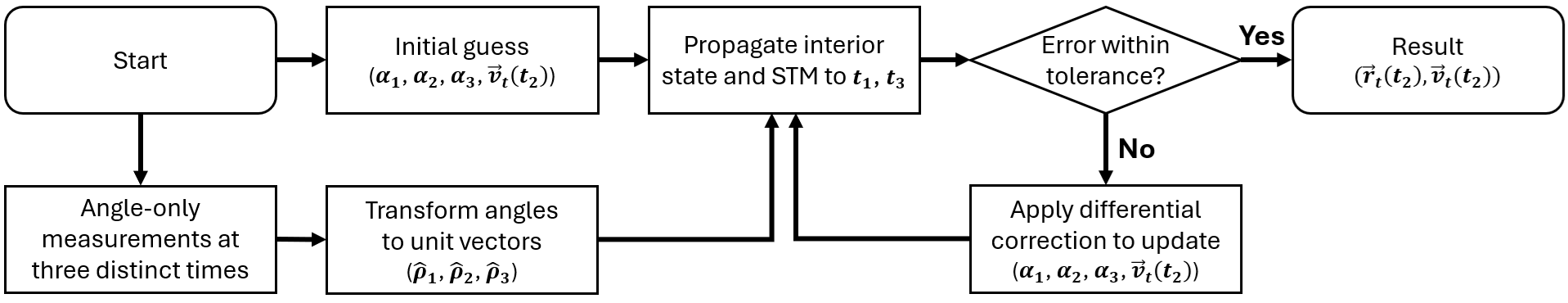}
    \caption{\label{fig:FlowChart} Flowchart of the IOD algorithm}
\end{figure}

Algorithm~\ref{alg:iod} and Fig.~\ref{fig:FlowChart} summarize the proposed angle-only IOD method in visual and procedural form. The algorithm estimates the target state at the interior measurement epoch using three LOS observations taken at distinct times. Each LOS is converted into a relative unit vector from the observer to the target. A Newton–Raphson scheme iteratively refines the range values and the associated velocity, while forward and backward propagation from the estimated state ensures consistency with all observed directions.

\begin{algorithm}[!htb]
\DontPrintSemicolon

\KwIn{Three angle-only measurements $\{a_k, h_k\}_{k=1}^{3}$ and observer positions $\vec{r}_o(t_k)$\;

\KwOut{Estimated target state $\vec{x}(t_2) = [\vec{r}_t(t_2);~\vec{v}_t(t_2)]$}

\BlankLine
Convert angles to unit vectors: $\hat{\rho}_k = R \cdot \begin{bmatrix} \cos h_k \cos a_k \\ \cos h_k \sin a_k \\ \sin h_k \end{bmatrix}$

\BlankLine
Initialize free variable vector: 
$\vec{\chi}_0 = \begin{bmatrix}
    \alpha_1^{(0)},~ \alpha_2^{(0)},~ \alpha_3^{(0)},~ \vec{v}_t^{(0)}(t_2)^\top
\end{bmatrix}^\top$
\hspace{1em}

\BlankLine
\While{$\|\vec{F}(\vec{\chi}^{(i)})\| >  \varepsilon$}{
    
    Compute target positions: $\vec{r}_t(t_k) = \vec{r}_o(t_k) + \alpha_k^{(i)} \hat{\rho}_k$
    
    \BlankLine
    Propagate $\vec{x}(t_2)$ to $t_1$ and $t_3$ $\Rightarrow$ $\tilde{r}_t(t_k)$
    
    \BlankLine
    Evaluate constraints:
    \[
    \vec{F}(\vec{\chi}^{(i)}) = \begin{bmatrix}
        \vec{r}_t(t_1) - \tilde{r}_t(t_1)\\
        \vec{r}_t(t_3) - \tilde{r}_t(t_3)
    \end{bmatrix}
    \]

    \BlankLine
    Compute Jacobian $DF(\vec{\chi}^{(i)})$:
    \[
    \begin{bmatrix} 
    \hat{\rho}_1 & -\Phi^-_{11}(t_1, t_2)\hat{\rho}_2 & 0_{3\times1} & -\Phi^-_{12}(t_1, t_2) \\
    0_{3\times1} & -\Phi^+_{11}(t_3, t_2)\hat{\rho}_2 & \hat{\rho}_3 & -\Phi^+_{12}(t_3, t_2) 
    \end{bmatrix}
    \]

    \BlankLine
    Update: $\vec{\chi}^{(i+1)} = \vec{\chi}^{(i)} - DF(\vec{\chi}^{(i)})^{-1} \vec{F}(\vec{\chi}^{(i)})$
    }
}
\caption{Angle-Only IOD via Differential Corrections}
\label{alg:iod}
\end{algorithm}

\section{Numerical Simulation} \label{sec:Simulation}

Numerical simulations are conducted across four distinct observer–target scenarios:
\begin{itemize}
\item \textbf{Scenario \#1:} A static observer at the lunar south pole performs IOD on a target in a southern $L_2$ 9:2 near-rectilinear halo orbit (NRHO).
\item \textbf{Scenario \#2:} An observer in an $L_5$ short-periodic orbit (SPO) estimates the state of a target approaching the $L_1$ gateway along its stable manifold.
\item \textbf{Scenario \#3:} A low-lunar orbit (LLO) observer tracks an NRHO target near perilune, capturing significant out-of-plane geometry.
\item \textbf{Scenario \#4:} An NRHO observer observes a target on an $L_1$ Lyapunov orbit.
\end{itemize}

\noindent Each scenario is evaluated using the proposed angle-only IOD algorithm. As described in Section~\ref{sec:Initialization}, the algorithm searches for a trajectory that satisfies the LOS constraints in the vicinity of the initial guess. Therefore, when multiple trajectories exist that fulfill the same constraints, the algorithm may converge to different solutions depending on the quality of the initial guess.

To characterize this sensitivity, the analysis performs a threshold-based evaluation of the initial guess quality required to consistently recover the true solution. For each scenario, a ground-truth set of range values $(\alpha_1^\ast, \alpha_2^\ast, \alpha_3^\ast)$ is defined, and the initial guesses $(\alpha_1^{(0)}, \alpha_2^{(0)}, \alpha_3^{(0)})$ are systematically varied. For simplicity and consistency, all three initial range guesses are set equal based on the second-epoch ground truth, such that $\alpha_1^{(0)} = \alpha_2^{(0)} = \alpha_3^{(0)} = \alpha_2^\ast + \epsilon$. The scalar $\epsilon$ represents the deviation from the true value and defines the convergence threshold within which convergence to the true solution is achieved, i.e., $\epsilon \in [\epsilon_{\min}, \epsilon_{\max}]$. This formulation captures potential differences in convergence behavior between underestimation and overestimation of the range values. The resulting convergence window, denoted by $(\epsilon_{\min}^{\text{conv}}, \epsilon_{\max}^{\text{conv}})$, thereby identifies the range of initial guesses that consistently lead to the true solution.

As an optional validation step, a fourth angle-only measurement can be incorporated to verify whether a converged solution corresponds to the correct physical trajectory, particularly in cases where ambiguity arises due to sensitivity to the initial guess or when additional confirmation is desired. Specifically, after performing IOD using measurements at $\{t_1, t_2, t_3\}$, the algorithm is re-run using $\{t_2, t_3, t_4\}$ with the previously estimated values of $\alpha_2$ and $\alpha_3$ serving as initial guesses. If the solutions coincide, the result is verified.

\subsection{Scenario \#1: Fixed Observer } \label{sec:sim1}
\begin{figure}[!htb]
\centering\includegraphics[width=0.8\linewidth]{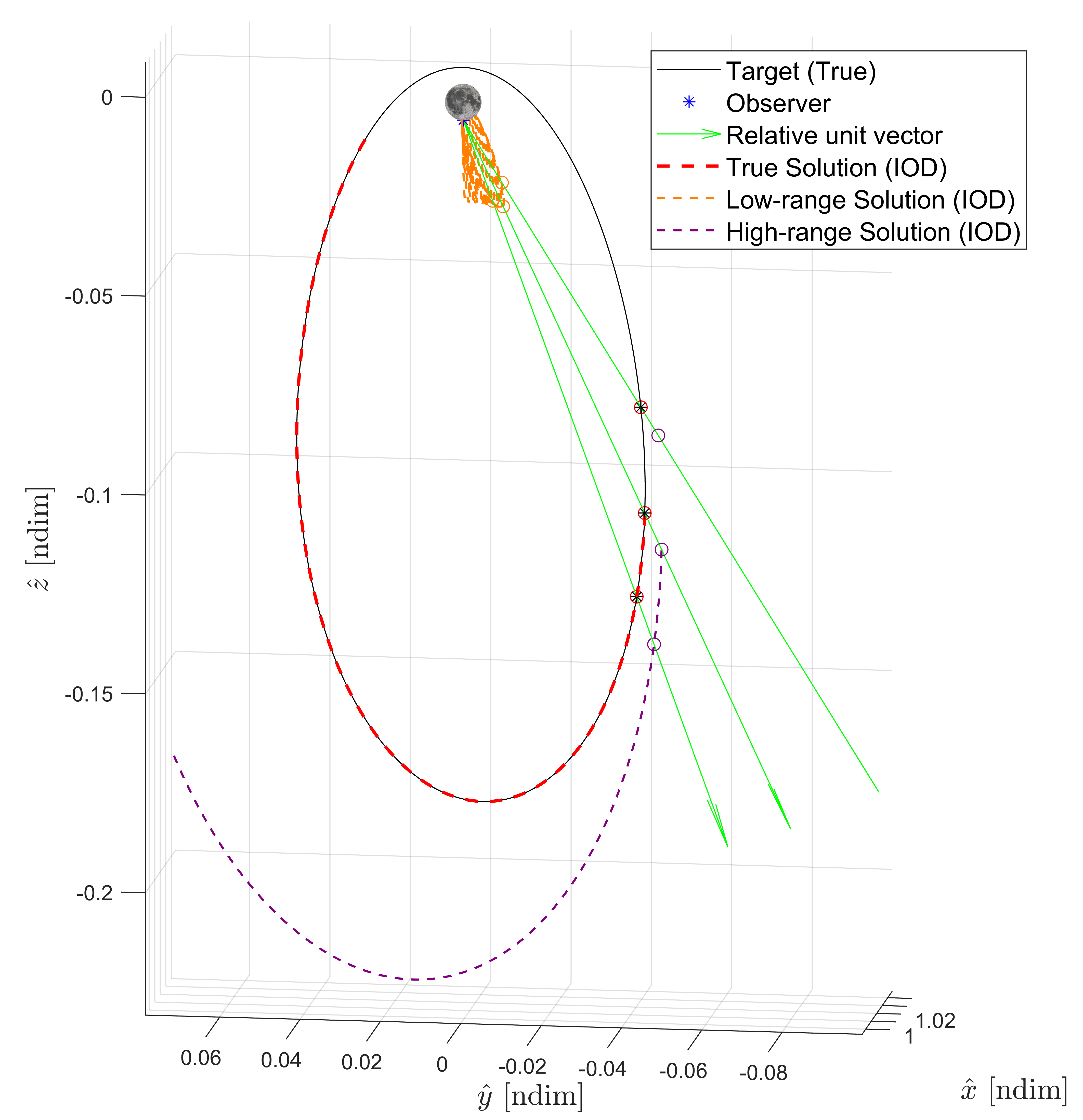}
	\caption{Estimated target trajectories under varying initial guesses (Scenario \#1)}
	\label{fig:solutions2}
\end{figure}

In this scenario, the observer remains stationary at the lunar south pole for the entire observation period and acquires measurements approximately every 7.98 hours on a target in an NRHO. Table~\ref{Scene1} summarizes the observer’s position, measurement times, and the corresponding relative unit vectors. For simplicity, the first measurement is taken at the start of the simulation.
\begin{table}[b!] 
\centering
\caption{System configuration in Scenario \#1.}
\label{Scene1}
\begin{tabular}{cccc}
\hline
             & Measurement 1             & Measurement 2             & Measurement 3             \\ \hline
Time {[}hrs{]}                       & 0                     & 7.9808                     & 15.9615                     \\ 
$\vec{r}_o\left(t\right)$ {[}km{]} & $\begin{bmatrix}379,\!729&0&-1,\!734\end{bmatrix}$ & $\begin{bmatrix}379,\!729&0&-1,\!734\end{bmatrix}$ & $\begin{bmatrix}379,\!729&0&-1,\!734\end{bmatrix}$ \\
$\hat{\rho}$                       & $\begin{bmatrix}0.1624&-0.3084&-0.9373\end{bmatrix}$ & $\begin{bmatrix}0.1486&-0.3871&-0.9100\end{bmatrix}$ & $\begin{bmatrix}0.1289&-0.4995&-0.8567\end{bmatrix}$ \\ \hline
\end{tabular}
\end{table}
The initial conditions are generated by perturbing the true range value at the second epoch ($\alpha_2^\ast$) by a scalar $\epsilon$, such that all three initial guesses are set equally as $\alpha_1^{(0)} = \alpha_2^{(0)} = \alpha_3^{(0)} = \alpha_2^\ast + \epsilon$. The algorithm successfully converges to the true solution when the common initial guess lies within the interval  $14,\!356~\text{km}\leq\alpha_1^{(0)}=\alpha_2^{(0)}=\alpha_3^{(0)}\leq44,\!010~\text{km} $ This corresponds to a convergence threshold for $\epsilon$ given by $\epsilon^{\text{conv}} \in [-28,\!265~\text{km}, ~+1,\!388~\text{km}]$, relative to the true value $\alpha_2^\ast$. For initial guesses outside this range, the algorithm tends to converge to either a low-range or high-range alternative solution, as summarized in Table~\ref{tab:case1sols}. The distribution of these solution groups is visualized in Figure~\ref{fig:solutions2}.

\begin{table}[!htb]
\centering
\caption{Converged IOD solutions (Scenario \#1)}
\label{tab:case1sols}
\begin{tabular}{cccc}
\hline
                        & $\alpha_1$ {[}km{]} & $\alpha_2$ {[}km{]} & $\alpha_3$ {[}km{]} \\ \hline
\textbf{\emph{Solution \#1 (True)}}       & \textbf{\emph{50,401}}               & \textbf{\emph{42,621}}               & \textbf{\emph{32,856}}               \\
Solution \#2 (Low-range)  & 8,532                & 9,371                & 7,161                \\
Solution \#3 (High-range) & 55,441               & 46,576               & 36,084               \\ \hline
\end{tabular}
\end{table}

\noindent As shown in Fig.~\ref{fig:solutions2}, among the three converged solutions, the Solution \#2 (Low-range) trajectory leads to a surface impact on the Moon, making it easily identifiable as physically invalid. However, distinguishing between the Solution \#3 (High-range) and the Solution \#1 (True) is more challenging, as both satisfy the LOS constraints without obvious physical inconsistencies. To resolve this ambiguity, if initial guess of $\alpha_2^{(0)}$ obtained from computer vision of images is uncertain that it lies within a valid range, the optional disambiguation strategy described in Section~\ref{sec:Initialization} is applied using a fourth angle-only measurement. Specifically, the estimated values of $\alpha_2$ and $\alpha_3$ obtained from the high-range solution are used as initial guesses, and the IOD is re-run using measurements at $\{t_2, t_3, t_4\}$. The algorithm then converges to the solution:
\begin{equation}
\alpha_2 = 42,\!621~\text{km}, \quad \alpha_3 = 32,\!856~\text{km}, \quad \alpha_4 = 20,\!129~\text{km},
\end{equation}
which matches the Solution \#1 (True) trajectory. This confirms that the Solution \#3 (High-range) obtained from the initial IOD is not physically valid, and the Solution \#1 (True) is correctly recovered through cross-validation.

\subsection{Scenario \#2: SPO Observer} \label{sec:sim2}
In this scenario, the observer follows an SPO near the $L_5$ point, with both the observer and the target remaining in-plane throughout the observation interval. The observer acquires angle-only measurements approximately every 28.69 hours. Table~\ref{Scene2} summarizes the observer’s positions and corresponding relative unit vectors at each measurement time. As in Scenario~\#1, the system is initialized at the time of the first measurement.

\begin{table}[h!] 
\centering
\caption{System configuration in Scenario \#2}
\label{Scene2}
\begin{tabular}{cccc}
\hline
             & Measurement 1             & Measurement 2             & Measurement 3             \\ \hline
Time {[}hrs{]}                       & 0                     & 28.6890                     & 57.3780                     \\ 
$\vec{r}_o\left(t\right)$ {[}km{]} & $\begin{bmatrix}-76,\!326&-321,\!899&0\end{bmatrix}$ & $\begin{bmatrix}-36,\!086&-298,\!824&0\end{bmatrix}$ & $\begin{bmatrix}18,\!145&-277,\!686&0\end{bmatrix}$ \\
$\hat{\rho}$                       & $\begin{bmatrix}-0.6040&0.7970&0\end{bmatrix}$ & $\begin{bmatrix}-0.6952&0.7188&0\end{bmatrix}$ & $\begin{bmatrix}-0.7881&0.6156&0\end{bmatrix}$ \\ \hline
\end{tabular}
\end{table}

\begin{figure}[!htb]
\centering
\includegraphics[width=0.85\textwidth]{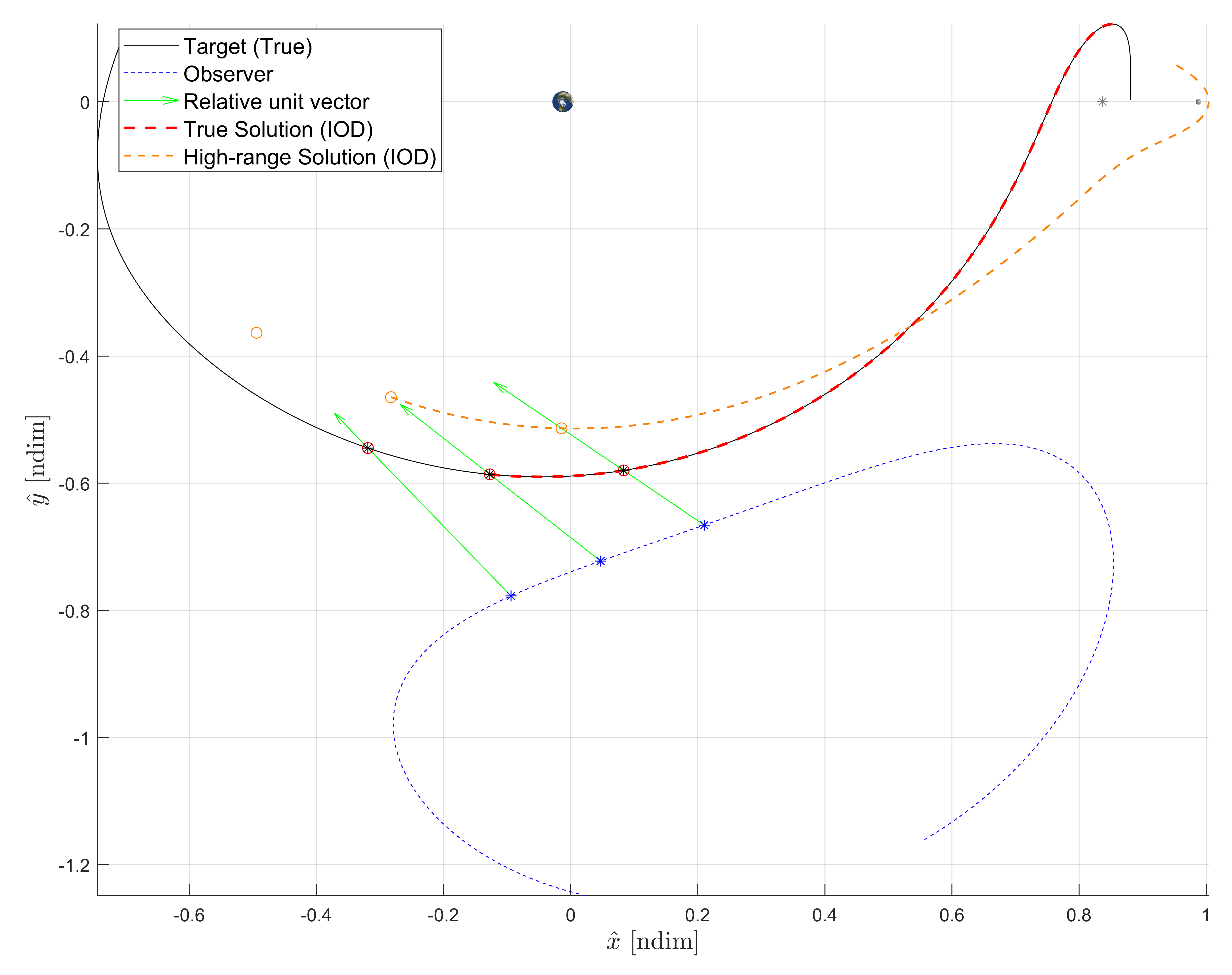}
\caption{Estimated target trajectories under varying initial guesses (Scenario \#2)}
\label{fig:solutions1}
\end{figure}

\noindent The initial conditions are generated by perturbing the true range value at the second epoch ($\alpha_2^\ast$) by a scalar $\epsilon$, such that all three initial guesses are set equally as
$\alpha_1^{(0)} = \alpha_2^{(0)} = \alpha_3^{(0)} = \alpha_2^\ast + \epsilon$. The algorithm successfully converges to the true solution when the common initial guess lies within the interval $49,\!972~\text{km} \leq \alpha_1^{(0)} = \alpha_2^{(0)} = \alpha_3^{(0)} \leq 142,\!228~\text{km}$.
This corresponds to a convergence threshold for $\epsilon$ given by
$\epsilon^{\text{conv}} \in [-35,\!147~\text{km}, ~+57,\!109~\text{km}]$, relative to the true value $\alpha_2^\ast$. For initial guesses outside this range, the algorithm converges to one of two alternative solutions: a low-range solution that aligns with the observer’s orbital path, or a high-range solution that overestimates the target distance. The former arises when the initial guess is significantly underestimated, causing the algorithm to interpret the measurements as originating from a co-located target. The latter results from an overestimated range and corresponds to a dynamically feasible but physically incorrect configuration. Each of these solution groups is summarized in Table~\ref{tab:case2sols} and visualized in Figure~\ref{fig:solutions1}.

\begin{table}[!htb]
\centering
\caption{Converged IOD solutions (Scenario \#2)}
\label{tab:case2sols}
\begin{tabular}{cccc}
\hline
                        & $\alpha_1$ {[}km{]} & $\alpha_2$ {[}km{]} & $\alpha_3$ {[}km{]} \\ \hline
\textbf{\emph{Solution \#1 (True)}}       & \textbf{\emph{124,412}}               & \textbf{\emph{85,119}}               & \textbf{\emph{58,892}}               \\
Solution \#2 (High-range) & 221,393               & 160,931               & 104,261              \\
\hline
\end{tabular}
\end{table}

\noindent As in Scenario~\#1, a fourth measurement is used to confirm the validity of a given solution. When the IOD is re-run using measurements at $\{t_2, t_3, t_4\}$ with the previously estimated values of $\alpha_2$ and $\alpha_3$ serving as initial guesses, the algorithm converges to the known true trajectory:
\begin{equation}
\alpha_2 = 85,\!119~\text{km}, \quad \alpha_3 = 58,\!892~\text{km}, \quad \alpha_4 = 47,\!941~\text{km}.
\end{equation}
This consistency verifies that the Solution \#1 (True) obtained using $\{t_1, t_2, t_3\}$ corresponds to the correct physical state. The fourth-measurement check thus provides a lightweight and reliable method to disambiguate between candidate solutions, especially in the presence of uncertain or ambiguous initialization.

\subsection{Scenario \#3: LLO Observer} \label{sec:sim3}
This scenario considers an observer in a LLO tracking a target near perilune. Compared to previous scenarios, measurements are taken at much shorter intervals, and both the observer and the target exhibit significant out-of-plane motion. This creates a high-dynamic environment with increased nonlinearities due to proximity to the Moon. The observer’s positions and relative unit vectors to the target at each measurement epoch are provided in Table~\ref{Scene3}.

\begin{table}[!htb] 
\centering
\caption{System configuration in Scenario \#3.}
\label{Scene3}
\begin{tabular}{cccc}
\hline
             & Measurement 1             & Measurement 2             & Measurement 3             \\ \hline
Time {[}mins{]}                       & 0                     & 9.5769                     & 19.1538                     \\ 
$\vec{r}_o\left(t\right)$ {[}km{]} & $\begin{bmatrix}
    379,\!729&-72&1,\!837
\end{bmatrix}$ & $\begin{bmatrix}379,\!735&835&1,\!638\end{bmatrix}$ & $\begin{bmatrix}379,\!742&1,\!529&1,\!022\end{bmatrix}$ \\
$\hat{\rho}$                       & $\begin{bmatrix}-0.1152&0.3004&0.9468\end{bmatrix}$ & $\begin{bmatrix}-0.1077&0.3066&0.9457\end{bmatrix}$ & $\begin{bmatrix}-0.0812&0.3404&0.9368\end{bmatrix}$ \\ \hline
\end{tabular}
\end{table}

\begin{figure}[!htb]
\centering\includegraphics[width=0.9\linewidth]{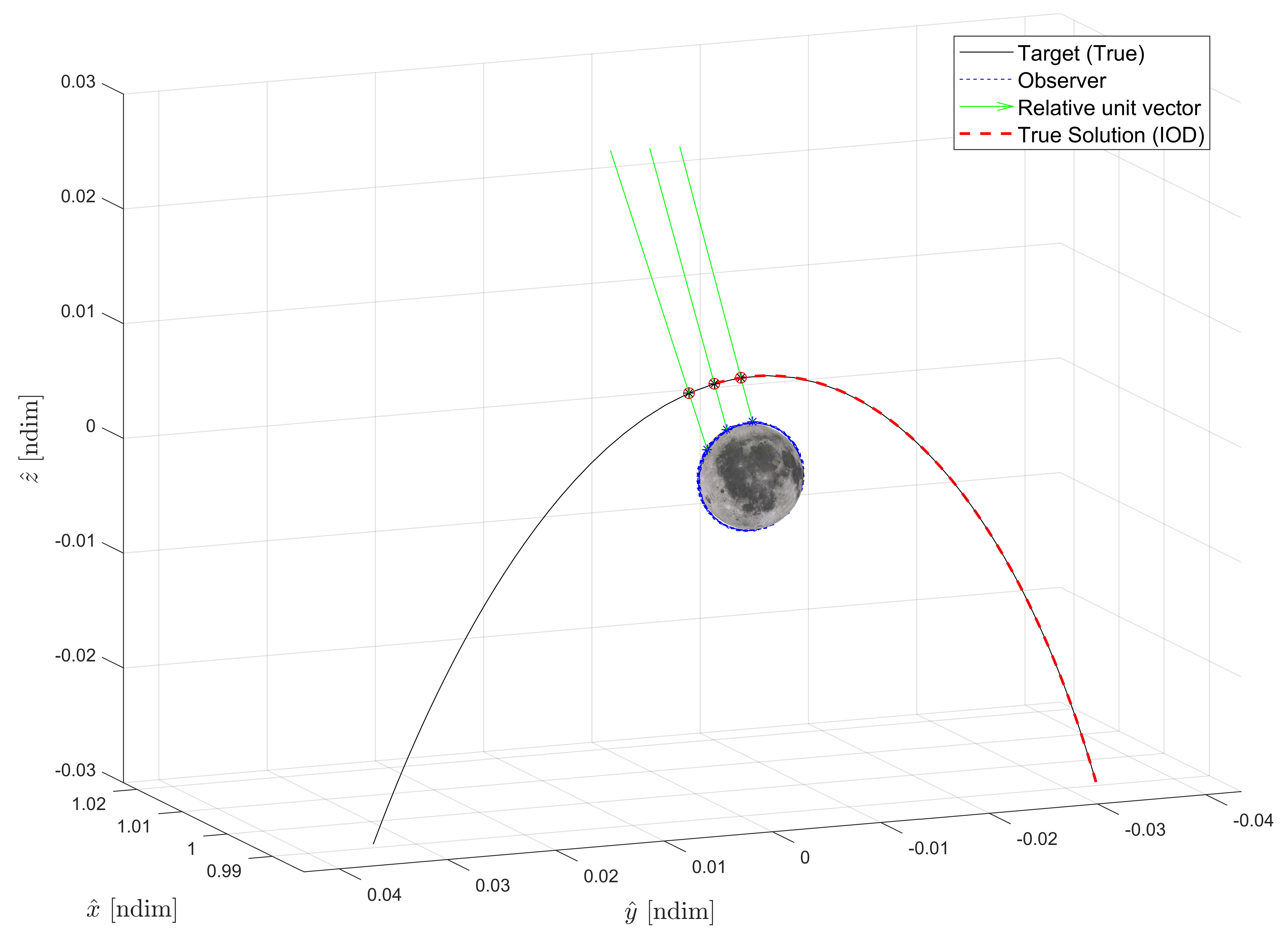}
\caption{Estimated target trajectory under varying initial guesses (Scenario \#3)}
\label{fig:solutions4}
\end{figure}

\noindent As shown in Fig.~\ref{fig:solutions4}, all trials consistently converge to a single, correct solution:
\begin{equation}
\alpha_1 = 1,\!633~\text{km},\quad \alpha_2 = 1,\!711~\text{km},\quad \alpha_3 = 2,\!122~\text{km}.
\end{equation}

\noindent In this scenario, despite the short observation intervals, close proximity, significant out-of-plane trajectories, and strong nonlinearities near the Moon, the algorithm demonstrates robust and rapid convergence to the true solution across all initial guesses.

\subsection{Scenario \#4: NRHO Observer} \label{sec:sim4}
This scenario evaluates the performance of the proposed IOD algorithm when the observer resides in an NRHO and tracks a target undergoing significant out-of-plane motion at a considerable distance. Compared to previous cases, the measurements are more widely spaced in time. Table~\ref{Scene4} summarizes the system configuration used in this scenario.

\begin{table}[h!] 
\centering
\caption{System configuration in Scenario \#4.}
\label{Scene4}
\begin{tabular}{cccc}
\hline
             & Measurement 1             & Measurement 2             & Measurement 3             \\ \hline
Time {[}hrs{]}                       & 0                     & 28.6890                     & 57.3780                     \\ 
$\vec{r}_o\left(t\right)$ {[}km{]} &$\begin{bmatrix}392,\!858&-4,\!893&-68,\!818\end{bmatrix}$ & $\begin{bmatrix}389,\!218&-14,\!146&-54,\!492\end{bmatrix}$ & $\begin{bmatrix}382,\!138&-14,\!883&-19,\!032\end{bmatrix}$ \\
$\hat{\rho}$                       & $\begin{bmatrix}-0.0912&0.9061&0.0413\end{bmatrix}$ & $\begin{bmatrix}0.0359&0.9267&0.3740\end{bmatrix}$ & $\begin{bmatrix}0.1765&0.9677&0.1799\end{bmatrix}$ \\ \hline
\end{tabular}
\end{table}

\begin{figure}[!b]
\centering\includegraphics[width=1\linewidth]{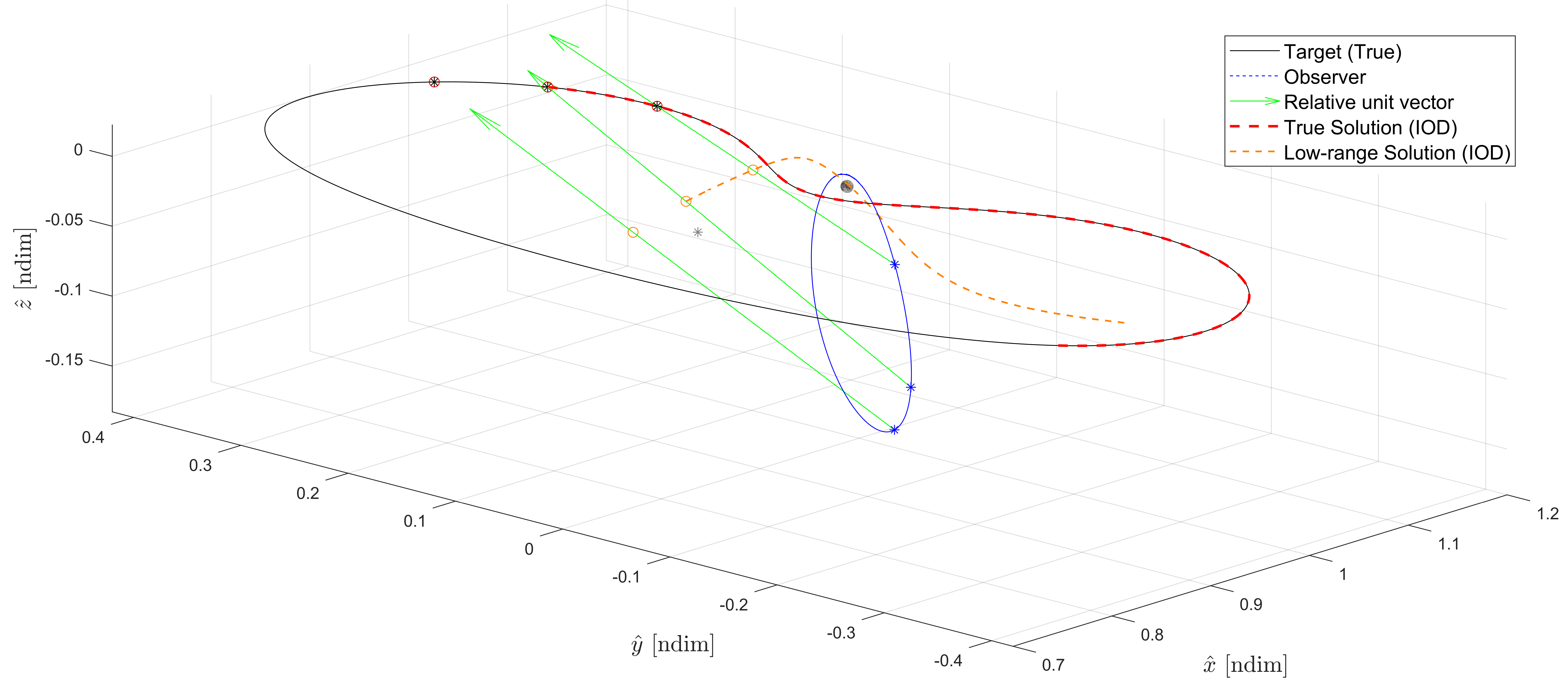}
\caption{Estimated target trajectories under varying initial guesses (Scenario \#4)}
\label{fig:solutions3}
\end{figure}

\noindent The initial conditions are generated by perturbing the true range value at the second epoch ($\alpha_2^\ast$) by a scalar $\epsilon$, such that all three initial guesses are set equally as
$\alpha_1^{(0)} = \alpha_2^{(0)} = \alpha_3^{(0)} = \alpha_2^\ast + \epsilon$. The algorithm successfully converges to the true solution when the common initial guess lies within the interval $
111,\!476~\text{km} \leq \alpha_1^{(0)} = \alpha_2^{(0)} = \alpha_3^{(0)}$. This corresponds to a convergence threshold for $\epsilon$ given by $\epsilon^{\text{conv}} \in [-34,\!222~\text{km},~ +\infty)$, relative to the true value $\alpha_2^\ast$. For smaller values, it converges to an incorrect low-range trajectory. Table~\ref{tab:case4sols} summarizes each solutions, and they are visualized in Fig.~\ref{fig:solutions3}.

\begin{table}[!htb]
\centering
\caption{Converged IOD solutions (Scenario \#4)}
\label{tab:case4sols}
\begin{tabular}{cccc}
\hline
                        & $\alpha_1$ {[}km{]} & $\alpha_2$ {[}km{]} & $\alpha_3$ {[}km{]} \\ \hline
\textbf{\emph{Solution \#1 (True)}}       & \textbf{\emph{166,583}}               & \textbf{\emph{145,698}}               & \textbf{\emph{105,807}}               \\
Solution \#2 (Low-range)  & 94,624                & 90,232                & 63,179                \\
\hline
\end{tabular}
\end{table}

\noindent As in previous scenarios, a fourth angle-only measurement may be optionally incorporated to verify whether the solution obtained from a given initial guess corresponds to the true trajectory. Specifically, when the algorithm is re-run using $\{t_2, t_3, t_4\}$ with the $\alpha_2$ and $\alpha_3$ values obtained from the Solution \#2 (Low-range) as initial guesses, it successfully converges to the solution:
\begin{equation} \label{eq:sol4}
\alpha_2 = 145,\!698~\text{km}, \quad \alpha_3 = 105,\!807~\text{km}, \quad \alpha_4 = 48,\!232~\text{km}.
\end{equation}
This result clearly validates that Solution \#1 (True), derived from ${t_1, t_2, t_3}$, corresponds to the true physical trajectory, as its $\alpha_2$ and $\alpha_3$ values match those obtained independently from ${t_2, t_3, t_4}$ in Eq.~\eqref{eq:sol4}.

\subsection{Summary of Simulations} \label{sec:sim_summary}
Across all scenarios, the proposed IOD method consistently recovers the true target state when the initial range guesses fall within a scenario-dependent threshold, denoted by $\epsilon^{\text{conv}}$, as summarized in Table~\ref{tab:summary}. These thresholds are computed relative to the second-epoch ground truth $\alpha_2^\ast$. The results demonstrate that convergence to the correct solution is achievable even in the absence of accurate range information. Notably, Scenario~\#3 exhibits global convergence across all initial guesses, while the other scenarios also show robust convergence behavior within reasonably wide thresholds. In contrast, Scenario~\#1 exhibits strong sensitivity to overestimation of the initial guess. Nevertheless, any ambiguity arising from such sensitivities can be resolved using the optional fourth-measurement strategy, confirming the robustness of the proposed algorithm under various geometric configurations.

\begin{table}[!htb]
\centering
\caption{Convergence results across all scenarios relative to the second-epoch range value, $\alpha_2$.}
\label{tab:summary}
\begin{tabular}{lccc}
\hline
\textbf{Scenario} & $\boldsymbol{\alpha_2^\ast}$ [km] & $\boldsymbol{\epsilon^{\text{conv}}}$ [km] & $\boldsymbol{\epsilon^{\text{conv}}}$ [\%] \\
\hline
Scenario \#1 (Fixed Observer)   & 42,621   & $[-28,\!265,~+1,\!388]$  & $[-66.3\%,~+3.3\%]$  \\
Scenario \#2 (SPO Observer)     & 124,412  & $[-35,\!147,~+57,\!109]$ & $[-28.2\%,~+45.9\%]$ \\
Scenario \#3 (LLO Observer)     & 1,711    & $(-\infty,~+\infty)$ & $(-\infty,~+\infty)$ \\
Scenario \#4 (NRHO Observer)    & 145,698  & $[-34,\!222,~+\infty)$ & $[-23.5\%,~+\infty)$ \\
\hline
\end{tabular}
\end{table}

\section{Conclusion} \label{sec:Conclusion}
This paper introduces a novel IOD algorithm for cislunar applications based solely on angle-only measurements. The core methodology is built upon a differential corrections framework that iteratively refines unknown range values to satisfy observed LOS constraints using the STM. While the formulation is well-constrained, it admits the possibility of multiple mathematically valid solutions that satisfy the same LOS constraints. To address this sensitivity, numerical simulations are conducted to empirically determine the convergence boundaries of the true solution for various scenarios. The results confirm that the proposed method consistently converges to the true solution across a reasonably wide range of initial guesses. An optional disambiguation strategy using a fourth angle-only measurement is also demonstrated, allowing the algorithm to reliably distinguish the true solution from alternative candidates when initialization uncertainty is present.

This structure enables fully autonomous operation without external intervention and enhances robustness in real-world deployment. The proposed method offers a practical, sensor-only solution for deep-space navigation, particularly in cislunar missions where Earth-based tracking support is limited or unavailable. In addition, the convergence time and number of iterations observed in the simulations demonstrate the algorithm’s efficiency and ease of implementation compared to alternative methods. Future work will focus on validating the algorithm using simulated imagery, extending it to account for sensor noise, and integrating it with onboard vision-based navigation systems for real-time operation.





\appendix

\subsection*{Acknowledgments}
\label{acknowledgements}
The authors acknowledge the help and support provided by the Space Trajectories and Applications Research (STAR) Lab at Embry-Riddle Aeronautical University.

\subsection*{Declaration of competing interest} 

The authors have no competing interests to declare that are relevant to the content of this article.

\section*{References}
\bibliographystyle{astrobib}
\bibliography{bib}

\subsection*{Author biography}

\begin{biography}[Seur]{Seur Gi Jo} is a Ph.D. student in Aerospace Engineering at Embry-Riddle Aeronautical University. His research focuses on understanding chaotic behavior in multi-body regimes through the application of knot theory and universal templates, as well as developing dynamics and control strategies for spacecraft rendezvous, proximity operations, and landing in the cislunar region. He graduated from the Republic of Korea Air Force Academy and has been serving in the Republic of Korea Air Force since 2010. Currently, he holds the rank of Major and has accumulated approximately 1,200 flight hours as an F-16 fighter pilot. (jos2@my.erau.edu)  
\end{biography}

\begin{biography}[Brian_Picture]{Brian Baker-McEvilly} is pursuing his PhD in Aerospace Engineering with a focus in dynamics and control at Embry-Riddle Aeronautical University, Florida. His current projects include constellation design in the Earth-Moon system to observe objects traversing Cislunar space and finding maneuvers that are able to avoid optical detection between key sections of Cislunar space. His other research interests include attitude dynamics, control, estimation, and trajectory design. Brian is the Technology and Resource Management Lead of STAR group. (bakermcb@my.erau.edu)
\end{biography}

\begin{biography}[image]{David Canales} is an Assistant Professor of Aerospace Engineering at Embry-Riddle Aeronautical University (ERAU), where he founded the Space Trajectories and Applications Research (STAR) group and serves as ERAU’s representative to the Universities Space Research Association (USRA). He earned BSc and MSc degrees in Aerospace Engineering from the Polytechnic University of Catalonia and an MSc in Astrophysics, Particle Physics, and Cosmology from the University of Barcelona. He later completed his Ph.D. in Astrodynamics and Space Applications at Purdue University under NAE member Dr. Kathleen C. Howell. His professional experience includes developing Earth-observation cameras for micro-satellites at Satlantis LLC. His research focuses on astrodynamics in multi-body regimes, cislunar operations, space weather, applied mathematics, and space policy. He was selected as a 2025 NASA PI Launchpad participant and is a 2025 Christine Mirzayan Science and Technology Policy Fellow with the National Academies’ Space Studies Board. (canaled4@erau.edu)
\end{biography}

\newpage
\vspace*{2.6em}
\subsection*{Graphical table of contents}

\begin{figure}[htb]
\centering
\includegraphics[width=150mm]{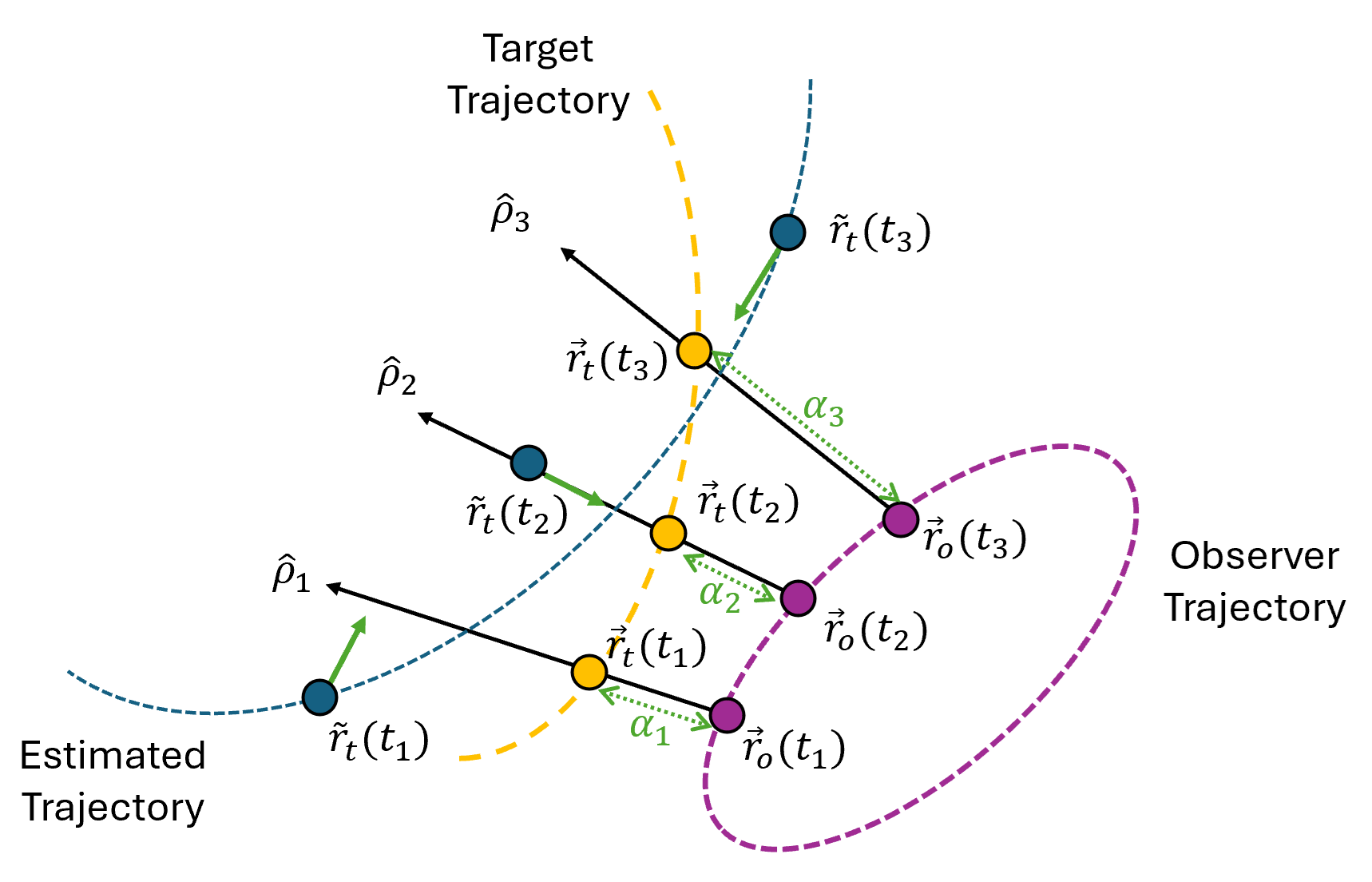}
\caption*{A novel angle-only initial orbit determination framework is proposed for the circular restricted three-body problem. Using only three line-of-sight measurements at distinct epochs, the full target trajectory is recovered via a differential corrections algorithm.}
\end{figure}

\end{document}